\documentclass{article}
\usepackage[margin=1in]{geometry}
\usepackage{amsthm,amsmath,amssymb}
\usepackage{graphicx}
\usepackage{natbib}
\usepackage[hidelinks,breaklinks]{hyperref}
\newtheorem{theorem}{Theorem}[]

\newtheorem{remark1}[theorem]{Remark}

\title{Ties in ranking scores can be treated as weighted samples}
\author{Mark Tygert}

\begin{document}

\maketitle

\begin{abstract}
Prior proposals for cumulative statistics suggest making
tiny random perturbations to the scores (independent variables in a regression)
in order to ensure the scores' uniqueness.
Uniqueness means that no score for any member of the population
or subpopulation being analyzed is exactly equal to any other member's score.
It turns out to be possible to construct from the original data
a weighted data set that modifies the scores, weights, and responses
(dependent variables in the regression) such that the new scores are unique
and (together with the new weights and responses)
yield the desired cumulative statistics for the original data.
This reduces the problem of analyzing data with scores that may not be unique
to the problem of analyzing a weighted data set with scores that are unique
by construction. Recent proposals for cumulative statistics
have already detailed how to process weighted samples whose scores are unique.
\end{abstract}

\section{Introduction}
\label{intro}

Recent work of~\cite{tygert_full} and others proposes cumulative statistics
for regression analysis and for assessing the calibration
of probabilistic predictions.
The data sets considered involve ranking ``scores,''
which are also known as ``independent variables'' or ``covariates''
in the case of regression and as ``predicted probabilities''
in the case of calibration.
When the scores need not be unique, that is,
when multiple individuals from the data may share exactly the same score,
\cite{tygert_full} recommends perturbing the scores
very slightly at random in order to break the degeneracy.
The present note proposes an alternative to breaking ties at random.
Namely, the present paper constructs a weighted data set based
on the original data such that the modified scores, responses, and weights
produce cumulative statistics entirely consistent with those
for the original data. The new scores are unique, so the earlier methods
of~\cite{tygert_full} et al.\ apply directly.

The formulation proposed below is merely an alternative
to the earlier proposals, not necessarily superior.
The new proposal requires no randomization of the data analysis,
unlike the earlier analyses.
The graphs of the earlier analyses directly displayed all members
of the original data set, omitting no one.
In contrast, for each score that multiple individuals share,
the graphs for the newly proposed analysis display
only the average of those multiple individuals' responses.
Nevertheless, the corresponding scalar summary statistics
have the same interpretations and asymptotic calibrations of P-values
given by~\cite{tygert_pvals} (under the hypotheses considered there).
Thus, the previous and new proposals have advantages and disadvantages relative
to each other (though none of the disadvantages is substantial, admittedly).
Both are good options to have available.

\section{Methods}
\label{methods}

This section proposes methods for analyzing data sets
consisting of ordered triples of scores, responses, and weights.
Scores are the ``independent variables'' or ``covariates''
and responses are the ``dependent variables'' or ``outcomes'' for a regression.
The weights pertain to weighted sampling.
The scores and weights will be viewed as given and fully specified, not random.
The responses will be viewed as random
and should be probabilistically independent from each other,
though not necessarily identically distributed.
The weights can be any strictly positive real numbers.
An ``unweighted'' sample would simply have the weights be uniform,
all equal to each other; so all analysis presented below is relevant
for both weighted and unweighted sampling.
The weights need not sum to 1 in the notation below,
as every use of the weights will normalize by their sum directly.

\cite{tygert_full} fully analyzed only data sets in which the scores
are real-valued and all unique:
\begin{equation}
\label{unique}
S_1 < S_2 < \dots < S_n,
\end{equation}
where the inequalities are all strict.
The present note considers the case in which each score $S_k$
may appear multiple times --- say $n_k$ times --- in the data set.
With this notation of $n_k$ specifying the degeneracy of score $S_k$,
we define $W_k$ to be the sum of all $n_k$ of the original weights
associated with score $S_k$; denoting the original weights
by $W_k^{(1)}$, $W_k^{(2)}$, \dots, $W_k^{(n_k)}$, we thus define
\begin{equation}
\label{newweight}
W_k = \sum_{j=1}^{n_k} W_k^{(j)}
\end{equation}
for $k = 1$, $2$, \dots, $n$.
We define $R_k$ to be the weighted average
of all $n_k$ of the original real-valued responses associated with score $S_k$;
denoting the original responses
by $R_k^{(1)}$, $R_k^{(2)}$, \dots, $R_k^{(n_k)}$, we thus define
\begin{equation}
R_k = \frac{\sum_{j=1}^{n_k} R_k^{(j)} \, W_k^{(j)}}
           {\sum_{j=1}^{n_k} W_k^{(j)}}
\end{equation}
for $k = 1$, $2$, \dots, $n$.
This yields a data set consisting of the weighted sample
$(S_k, R_k, W_k)$ for $k = 1$, $2$, \dots, $n$,
where $S_k$ is the score, $R_k$ is the associated response,
and $W_k$ is the associated weight.
So this new weighted data set contains $n$ members
$(S_k, R_k, W_k)$ for $k = 1$, $2$, \dots, $n$,
whereas the original data set contains $\sum_{k=1}^n n_k$ members
$(S_k^{(j)}, R_k^{(j)}, W_k^{(j)})$ for $k = 1$, $2$, \dots, $n$;
$j = 1$, $2$, \dots, $n_k$.
Analyzing the new weighted data set via the cumulative statistics
is a good way to analyze the original data set.
And, unlike the scores for the original data set,
the scores for the new weighted data set are guaranteed to be unique.

\cite{tygert_full} proposed cumulative statistics for analyzing calibration
and for analyzing deviation of a subpopulation from the full population,
based on data consisting of scores, responses, and weights. We now show that
the cumulative statistics for the original and new data sets are consistent
with each other.

The cumulative differences for the new data are
\begin{equation}
\label{newcumdiffs}
C_{\ell} = \frac{\sum_{k=1}^{\ell} (R_k - r(S_k)) \, W_k}{\sum_{k=1}^n W_k}
\end{equation}
for $\ell = 1$, $2$, \dots, $n$,
where $r$ is the regression function we seek to test;
when testing calibration, the regression function $r$
is simply the identity function $r(s) = s$ for every real number $s$.
When comparing a subpopulation to the full population,
$r(S_k)$ would be the (weighted) average of responses
from the full population at scores that are closer to $S_k$
than to any other of the scores $S_1$, $S_2$, \dots, $S_n$.
We set $C_0 = 0$, too. 

Let us denote by $v(R_k)$ the variance of the response $R_k$
corresponding to the score $S_k$ under the null hypothesis,
where the null hypothesis makes assumptions about the original data directly
(so that inferences about $R_k$ take into account the fact that $R_k$
is a weighted average of other random variables, instead of considering $R_k$
to be a single response variable).
For example, under the null hypothesis of perfect calibration
with each response drawn independently from a Bernoulli distribution,
\begin{equation}
\label{Bernoulli}
v(R_k) = S_k \, (1 - S_k) \, \frac{\sum_{j=1}^{n_k} \left(W_k^{(j)}\right)^2}
                                  {\left(\sum_{j=1}^{n_k} W_k^{(j)}\right)^2},
\end{equation}
since $S_k \, (1 - S_k)$ is the variance
of the Bernoulli distribution whose expected value is $r(S_k) = S_k$.
Calibration need not be the only hypothesis of interest to test.
Under the null hypothesis that a subpopulation being assessed does not deviate
from the function $r$ for the full population,
an estimate of $v(R_k)$ can be the (weighted) average of variances of responses
from the full population at scores that are closer to $S_k$
than to any other of the scores $S_1$, $S_2$, \dots, $S_n$
(assuming as always that the responses are all independent),
multiplied by the same factor from~(\ref{Bernoulli}), namely
\begin{equation}
\label{factor}
\frac{\sum_{j=1}^{n_k} \left(W_k^{(j)}\right)^2}
     {\left(\sum_{j=1}^{n_k} W_k^{(j)}\right)^2};
\end{equation}
indeed, the independence of all the responses yields that $v(R_k)$
is equal to the quantity in~(\ref{factor}) times the variance of $R_k^{(j)}$,
for every $j = 1$, $2$, \dots, $n_k$; $k = 1$, $2$, \dots, $n$.
\cite{tygert_full} gives the details.
Since we assumed that the responses are independent, the variance
of $C_{\ell}$ from~(\ref{newcumdiffs}) under the null hypothesis is
\begin{equation}
\label{total}
(\sigma_{\ell})^2 = \frac{\sum_{k=1}^{\ell} v(R_k) \, (W_k)^2}
                         {(\sum_{k=1}^n W_k)^2}
\end{equation}
for $\ell = 1$, $2$, \dots, $n$.

We also consider similar cumulative differences for the original data set
in which the scores are perturbed infinitesimally at random
(so that the scores become unique):
\begin{equation}
\label{cumdiffs}
B_{\ell} = \frac{\sum_{k=1}^{\ell} \sum_{j=1}^{n_k}
                 \left(R_k^{(j)} - r(S_k)\right) \, W_k^{(j)}}
                {\sum_{k=1}^n \sum_{j=1}^{n_k} W_k^{(j)}}
         = \frac{\sum_{k=1}^{\ell} (R_k - r(S_k)) \sum_{j=1}^{n_k} W_k^{(j)}}
                {\sum_{k=1}^n \sum_{j=1}^{n_k} W_k^{(j)}}
\end{equation}
for $\ell = 1$, $2$, \dots, $n$,
where the ordering of $R_k^{(1)}$, $R_k^{(2)}$, \dots, $R_k^{(n_k)}$
(and the corresponding weights) is randomized
for each $k = 1$, $2$, \dots, $n$. We set $B_0 = 0$, too.

We define abscissae via the aggregations
\begin{equation}
A_{\ell} = \frac{\sum_{k=1}^{\ell} \sum_{j=1}^{n_k} W_k^{(j)}}
                {\sum_{k=1}^n \sum_{j=1}^{n_k} W_k^{(j)}}
         = \frac{\sum_{k=1}^{\ell} W_k}{\sum_{k=1}^n W_k}
\end{equation}
for $\ell = 1$, $2$, \dots, $n$, where the latter equality
follows from~(\ref{newweight}). We set $A_0 = 0$, too.
Combining~(\ref{newweight}), (\ref{newcumdiffs}), and~(\ref{cumdiffs})
shows that $B_{\ell} = C_{\ell}$
for all $\ell = 1$, $2$, \dots, $n$.
Therefore, the piecewise linear graph connecting
the points $(A_{\ell},\,B_{\ell} / \sigma_n)$
for $\ell = 0$, $1$, $2$, \dots, $n$
and the piecewise linear graph connecting
the points $(A_{\ell},\,C_{\ell} / \sigma_n)$
for $\ell = 0$, $1$, $2$, \dots, $n$ are the same.
This demonstrates that the cumulative statistics for the original
and new data sets are consistent with each other.
Indeed, the corresponding graph of cumulative differences
for the original data with its scores perturbed very slightly
(so that the scores become unique) is the same
aside from the other graphs linearly interpolating
from each score $S_k$ to the next greatest score, $S_{k+1}$,
rather than interpolating linearly from each and every perturbed score
to the next greatest perturbed score.

\section{Conclusion}
\label{conclusion}

The cumulative statistics of~\cite{tygert_full} for the original data set
can require minute random perturbations to the scores,
unlike the cumulative statistics for the new weighted data.
The randomization does preserve more information about the original data,
as the associated graph of cumulative differences displays
the response of every single individual from the original data set.
The new weighted data set instead avoids any randomization
but, for each score that multiple members share,
averages together the multiple members' responses.
Thus both the previous approaches and the new proposal have pros and cons
relative to each other.
That said, the approaches are more similar than different;
neither has any substantial drawback.

\section*{Acknowledgements}

We would like to thank Imanol Arrieta Ibarra, Kamalika Chaudhuri,
Hannah Korevaar, and Mike Rabbat for their interest, suggestions, and support.

\bibliography{degenerate}
\bibliographystyle{authordate1}

\end{document}